\documentclass{article}

\usepackage{array}
\usepackage{graphicx}
\usepackage{verbatim}
\usepackage{color}
\usepackage{amsmath}
\usepackage{hyperref}
\usepackage[affil-it]{authblk}
\usepackage{cite}
\newcommand{\ket}[1]{\vert{#1}\rangle}

\begin{document}

\title{Distillation protocols for Fourier states in quantum computing}
\author{Cody Jones%
  \thanks{Electronic address: \href{mailto:ncodyjones@gmail.com}{ncodyjones@gmail.com}}}
\affil{Edward L. Ginzton Laboratory, Stanford University, Stanford, California 94305-4088, USA}

\maketitle

\begin{abstract}
Fourier states are multi-qubit registers that facilitate phase rotations in fault-tolerant quantum computing.  We propose distillation protocols for constructing the fundamental, $n$-qubit Fourier state with error $O(2^{-n})$ at a cost of $O(n \log n)$ Toffoli gates and Clifford gates, or any arbitrary Fourier state using $O(n^2)$ gates.  We analyze these protocols with methods from digital signal processing.  These results suggest that phase kickback, which uses Fourier states, could be the current lowest-overhead method for generating arbitrary phase rotations.
\end{abstract}



\section{Introduction}
Fault-tolerant quantum computing seeks to develop quantum information processors that are resilient to faults in any hardware component using quantum error-correction codes~\cite{Preskill1998,Nielsen2000}.  Recent attention has focused on how to minimize the resource costs of essential quantum computing primitives, such as a ``universal'' set of operations.  Theoretical analysis has shown that there must be at least one operation required for universality but not natively supported by the quantum code (often called ``non-transversal'')~\cite{Zeng2007,Eastin2009}.  Under realistic device parameters, the resource costs of the creating non-transversal gates dominate the total overhead for achieving fault tolerance~\cite{Isailovic2008,Jones2012,Fowler2012,Fowler2012b,Fowler2013}.  Therefore, choosing the appropriate non-transversal operations directly impacts resource costs for fault-tolerant quantum computers.

Many investigations, including some involving this author, adopt a ``simpler is better'' approach to selecting which non-transversal gate completes the universal set, so they focus on the single-qubit gate $T = \exp[i\pi (I-\sigma^z)/8]$, where $I$ is identity~\cite{Bravyi2005,Dawson2005,Isailovic2008,Fowler2009,Clark2009,Jones2012,Jones2012b,Meier2012,Amy2012,Fowler2012,Fowler2012b,Selinger2012,Bravyi2012,Jones2012c,Kliuchnikov2012,Fowler2013,Amy2013}.  Recently, Landahl and Cesare proposed using a family of rotation gates that they label $Z_k = \exp[i \pi \phi (I - \sigma^z)/2^{k+1}]$, which includes $Z_2 = T$, in what they term ``complex instruction set computing (CISC)''~\cite{Landahl2013}; a similar approach was used as a subroutine in Ref.~\cite{Isailovic2008}.  Instead of the $T$~gate, we promote an alternative, the three-qubit Toffoli gate~\cite{Barenco1995,Nielsen2000,Jones2013,Eastin2012}.  As an aside, our results also allow one to implement the CISC proposal efficiently.  Ultimately, the universal gate set is used to approximate quantum circuits needed in an algorithm.  Toffoli gates are already preferred for arithmetic circuits~\cite{Cuccaro2004,VanMeter2005,Draper2006,Jones2012}, but this work goes further to show that Toffoli is also efficient for arbitrary gates.

Recent work has shown that any arbitrary gate can be efficiently approximated using an instruction set that includes the $T$~gate~\cite{Fowler2009,Kliuchnikov2012}.  It has also been shown that arbitrary gates can be efficiently approximated with phase kickback, which uses Toffoli gates, so long as one has a multi-qubit resource that we call a ``Fourier state''~\cite{Jones2012b}.  This paper completes the phase kickback method by showing how to construct the Fourier state efficiently.  Moreover, we go further to argue that phase kickback can have lower resource costs than methods that use $T$~gates; when this is true, one should consider using Toffoli as the non-transversal operation in fault-tolerant quantum computing.

To give some context, a Fourier state of size $n$ qubits is defined as
\begin{equation}
\ket{\gamma^{(k)}} = \frac{1}{\sqrt{N}}\sum_{y=0}^{N-1} e^{i 2 \pi ky/N}\ket{y},
\label{define_Fourier_state}
\end{equation}
where $N = 2^n$.  Note that this is the sign convention of Ref.~\cite{Nielsen2000} and opposite of that in Ref.~\cite{Jones2012b}.  These states are eigenstates of the modular addition operator $U_{\oplus 1}\ket{x} = \ket{x + 1 \; (\mathrm{mod} \; N)}$:
\begin{equation}
U_{\oplus 1}\ket{\gamma^{(k)}} = e^{-i 2 \pi k/N}\ket{\gamma^{(k)}}.
\end{equation}
Using this property, a phase-rotation gate can be approximated using a Fourier state and an addition circuit, which is known as phase kickback~\cite{Cleve1998,Kitaev2002,Jones2012b}.  This method can produce any rotation around the $\sigma^z$ axis of the Bloch sphere in units of $\pi/2^{n-1}$ radians, so the precision required by a quantum algorithm determines $n$.  A notable feature of phase kickback is that the register $\ket{\gamma^{(k)}}$ is preserved, which means it can be used repeatedly.  Many implementations of addition circuits are known~\cite{Cuccaro2004,VanMeter2005,Draper2006}, but the fault-tolerant preparation of Fourier states has received less attention.  Kitaev \emph{et~al.} propose a scheme based on phase estimation~\cite{Kitaev2002}, but this protocol suffers from two notable disadvantages.  First, the resulting state $\ket{\gamma^{(k)}}$ has random odd $k$.  Second, the protocol requires a Fourier transform, which requires phase rotations; since the purpose of phase kickback is to produce phase rotations, implementing the Fourier transform to produce the Fourier state requires an approximate, iterative procedure.  This paper develops a fault-tolerant distillation protocol for producing the frequently used $\ket{\gamma^{(1)}}$ state with $O(n \log n)$ gates from a finite set.  We also give a related protocol for constructing any $\ket{\gamma^{(k)}}$ using $O(n^2)$ gates.

The circuit for constructing Fourier states is implemented in a fault-tolerant quantum computer that, owing to the constraints of error correction~\cite{Preskill1998,Zeng2007,Eastin2009}, has a limited set of gates.  In what follows, we will denote Pauli operators by $Z \equiv \sigma^z$, \emph{etc}.  Some gates require more resource overhead to produce than others.  Gates in the Clifford group are generated by combinations of: Hadamard $H = (1/\sqrt{2})(X + Z)$, phase gate $S = \exp[i\pi(I - Z)/4]$, and CNOT, up to global phase that we ignore.  In the set of ``Clifford gates,'' we also include ancilla qubits intialized to $\ket{0}$ and measurement $M_z$ performed in the $Z$ basis.  We assume that any Clifford gate is economical in terms of resource cost and that non-Clifford gates (those outside the Clifford group) are the dominant cost.  This assumption is justified by analysis showing that non-Clifford gates are substantially more resource intensive than Clifford gates~\cite{Isailovic2008,Jones2012,Fowler2012,Fowler2012b,Fowler2013}.  Nevertheless, a universal set of operations requires at least one non-Clifford gate, and we select Toffoli.  Recent work shows that low-overhead constructions for the Toffoli gate exist, often by implementing low-fidelity $T$~gates with subsequent error correction~\cite{Jones2013,Eastin2012}.

The recent analysis of resource costs in Refs.~\cite{Fowler2012,Fowler2012b,Jones2013,Fowler2013} indicates that producing a Toffoli gate with error $10^{-12}$ requires comparable physical resources when using surface code error correction to producing a single $T$~gate having the same error probability.  Although Toffoli gates could be the more resource-efficient choice of non-Clifford gate, constructing quantum algorithms efficiently with Toffoli gates was not fully resolved (prior to this work) in situations where arbitrary phase rotations are required.  A major consequence of this paper is that it completes the phase-kickback protocol by showing that the discrete set of Clifford operations and Toffoli gates can efficiently approximate any quantum unitary, which could replace constructions using $T$~gates~\cite{Dawson2005,Fowler2009,Amy2012,Kliuchnikov2012}.

The paper is organized as follows.  Section~\ref{fundamental_distillation} presents a protocol for distilling the ``fundamental'' Fourier state $\ket{\gamma^{(1)}}$ from approximations produced using only Clifford gates.  Section~\ref{fundamental_analysis} analyzes the resource costs of the distillation protocol, which we summarize here.  Constructing an $n$-qubit $\ket{\gamma^{(1)}}$ state requires circuit width $2n + O(1)$ qubits, circuit depth $O(n)$ gates, and $O(n \log n)$ Toffoli gates in total.  Section~\ref{resource_comparison} compares phase kickback to a competing method using $T$~gates, which is relevant since the purpose of Fourier distillation is to complete fault-tolerant constructions for phase kickback.  Section~\ref{arbitrary_distillation} outlines a protocol for distilling $\ket{\gamma^{(k)}}$ with arbitrary $k$ using $O(n^2)$ Toffoli gates.  The paper concludes with a discussion of why the combination of these results and recent work in constructing Toffoli gates makes phase kickback a compelling approach to approximating arbitrary quantum gates with low overhead.

\section{Distilling the fundamental Fourier state}
\label{fundamental_distillation}
The fundamental $\ket{\gamma^{(1)}}$ state of size $n$ qubits is required for phase-kickback constructions for both single-qubit phase rotations and two-qubit, controlled phase rotations~\cite{Jones2012b}.  This state is also useful for constructing a quantum Fourier transform (or its approximate version) through a special form of phase kickback called quantum-variable rotation~\cite{Jones2012b}.  This section shows how to construct $\ket{\gamma^{(1)}}$ using a distillation protocol based on addition circuits.  We generalize the method in a later section to distill $\ket{\gamma^{(k)}}$ for arbitrary $k$, but the special case of $k = 1$ requires fewer quantum gates.

\subsection{Approximate fundamental Fourier state}
Any pure Fourier-basis state where $N$ in Eqn.~(\ref{define_Fourier_state}) is a power of 2 is separable into individual qubits.  Using a general $Z$-axis rotation $R_z(\phi) = \exp[i \pi \phi (I - Z)/2]$, a Fourier state can be decomposed as
\begin{eqnarray}
\ket{\gamma^{(k)}} = \left[ R_z(k\pi/2^0)\ket{+} \right] \otimes \left[ R_z(k\pi/2^1)\ket{+} \right] \otimes \left[ R_z(k\pi/2^2)\ket{+} \right] \otimes (\ldots).
\label{Fourier_state_decomposition}
\end{eqnarray}
The single-qubit state $\ket{+} = H \ket{0} = (1/\sqrt{2})(\ket{0} + \ket{1})$.  Using Eqn.~(\ref{Fourier_state_decomposition}), we can see that the quantum state
\begin{equation}
\ket{\tilde{\gamma}^{(1)}} = Z\ket{+} \otimes S\ket{+} \otimes I\ket{+} \otimes I\ket{+} \otimes (\ldots)
\label{initial_state}
\end{equation}
is an approximation of $\ket{\gamma^{(1)}}$ (denoted with tilde).  Moreover, $\ket{\tilde{\gamma}^{(1)}}$ can be produced using only Clifford gates.  The rotations of $\pi/4$, $\pi/8$, \emph{etc.} in Eqn.~(\ref{Fourier_state_decomposition}) that are omitted in Eqn.~(\ref{initial_state}) become exponentially close to identity (in gate fidelity) with increasing qubit index, so we approximate them with identity gates.  This is the same justification behind neglecting small-angle rotations in the approximate quantum Fourier transform~\cite{Barenco1996}.  The fidelity between the approximate and ideal states is $\left|\left\langle \tilde{\gamma}^{(1)} \mid \gamma^{(1)} \right\rangle\right|^2 \ge 0.81$ for all values of $n$.

Each approximate state can be expanded in the orthonormal Fourier-state basis as
\begin{equation}
\ket{\tilde{\gamma}^{(1)}} = \sum_{j = 0}^{N-1} a_j \ket{\gamma^{(j)}},
\label{Fourier_basis}
\end{equation}
where the dominant term among the complex coefficients is $a_1$, with magnitude $\left|a_1\right|^2 \approx 0.81$ from above.  We ignore complex phase because we always work in the Fourier basis, and our distillation protocol depends only on the magnitudes of the Fourier-basis coefficients.

\subsection{Distillation protocol}
\label{protocol_section}
Using two approximate $\ket{\tilde{\gamma}^{(1)}}$ states, the distillation protocol is very simple.  First, add one register to the other.  Binary-encoded, mod-$2^n$ addition given by
\begin{equation}
U_{\mathrm{add}}\ket{v}\ket{w} = \ket{v}\ket{w+v \; (\mathrm{mod} \; 2^n)}
\end{equation}
has been studied extensively~\cite{Cuccaro2004,VanMeter2005,Draper2006}.  Notably, many addition circuits use the Toffoli gate as the non-Clifford operation.  In the Fourier basis, the addition circuit implements
\begin{equation}
U_{\mathrm{add}}\ket{\gamma^{(k)}}\ket{\gamma^{(k')}} = \ket{\gamma^{(k-k')}}\ket{\gamma^{(k')}}.
\end{equation}
As an aside, this is precisely phase kickback, where the Fourier index $k'$ of the second register determines the quantum-variable rotation applied to the first register (see Section~4.1 of Ref.~\cite{Jones2012b}).  Second, measure the first register in the Fourier basis, and postselect the cases where the result is $\ket{\gamma^{(0)}}$.  The resulting output has each of its Fourier-basis coefficients $\{a_k\}$ weighted by the probability that \emph{both} inputs to distillation were in the $\ket{\gamma^{(k)}}$ state.  If both inputs had sizable overlap with a particular state, then the fidelity conditioned on successful distillation is concentrated to a higher magnitude, and probability of being in unwanted basis states is suppressed.

\begin{figure}
  \centering
  \includegraphics[width=10cm]{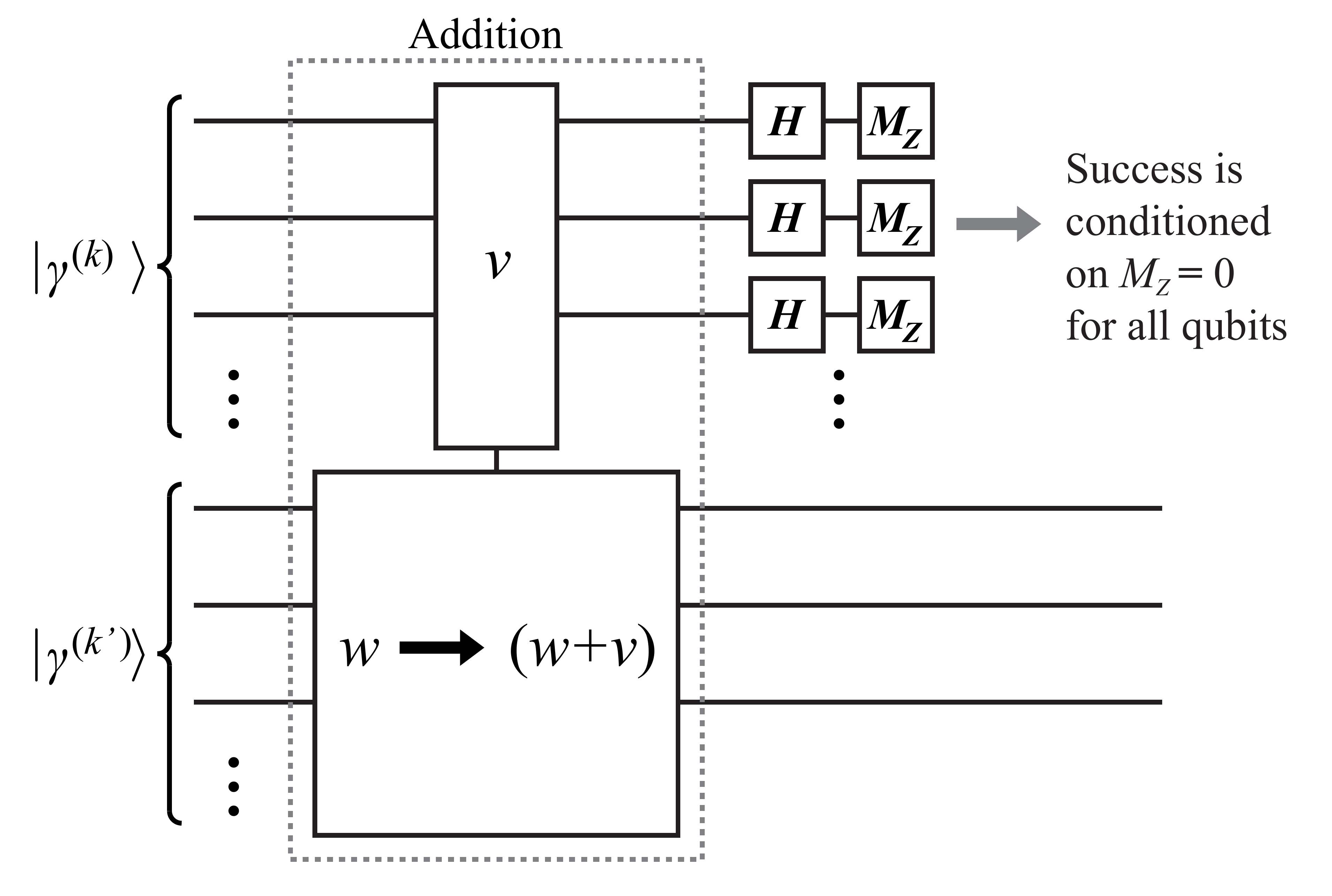}\\
  \caption{Circuit for distilling Fourier states.  Each Fourier state is a multi-qubit register, and only three qubits are shown, with the rest indicated by vertical dots.  For clarity, this circuit shows the inputs as pure Fourier-basis states, but in actual distillation protocols, the inputs will be mixed states.  The addition circuit shown in the dashed box would be decomposed into Clifford gates and Toffoli gates. The verification step is implemented with Hadamard $H$ and computational-basis measurement $M_z$ on each qubit in the top register.}
  \label{distillation_circuit}
\end{figure}

The quantitative expressions for distillation success probability and projection of output state into the Fourier basis are simple to derive.  Let the two inputs to distillation have Fourier-basis coefficients $\{a_j\}$ and $\{a_j'\}$ as in Eqn.~(\ref{Fourier_basis}).  The probability of measuring $\ket{\gamma^{(0)}}$ (\emph{i.e.} distillation succeeds) is given by
\begin{equation}
P_{\mathrm{success}} = \sum_{y = 0}^{N-1}\left|a_y\right|^2\left|a_y'\right|^2.
\end{equation}
The output register of distillation will have Fourier coefficients $\{b_j\}$ with magnitudes
\begin{equation}
\left|b_j\right|^2 = \frac{\left|a_j\right|^2\left|a_j'\right|^2}{P_{\mathrm{success}}}.
\end{equation}
These expressions mirror those of entanglement distillation~\cite{Bennett1996,Deutsch1996}.  We measure the fidelity of the output state as $F = \left|b_1\right|^2$, and the error probability in the distilled state is $\epsilon = 1-F$.

A general Fourier-basis measurement would pose a problem because it requires operations outside the Clifford group, but we show how to circumvent this issue with Clifford gates.  Since the quantum computer only supports computational-basis measurements, we would require a quantum Fourier transform (QFT) to map between the bases.  This is essentially the obstacle encountered by the Kitaev-Shen-Vyalyi protocol~\cite{Kitaev2002}, which addresses the matter with an iterative procedure of approximate QFTs.  However, our distillation protocol does not require a \emph{complete} Fourier-basis measurement; instead, we only need to know if the first register is in state $\ket{\gamma^{(0)}}$.  This state happens to be the tensor product of $\ket{+}$ states, which are eigenstates in the $X$-basis.  Hence, we only require application of the Hadamard gate $H$ followed by measurement $M_z$ on each qubit in the first register in Fig.~\ref{distillation_circuit}.  If each qubit is the $\ket{+}$ state, then the register was projected into $\ket{\gamma^{(0)}}$, and distillation succeeds.  Otherwise, reject the output and attempt again.  Since $H$ and $M_z$ are in the Clifford group, they are considered inexpensive to produce relative to the preceding non-Clifford addition circuit.  In addition to preparing the specific state $\ket{\gamma^{(1)}}$ (as opposed to a random Fourier-basis state~\cite{Kitaev2002,Jones2012b}), this measurement trick is how our protocol improves on the method in Ref.~\cite{Kitaev2002}.

If the two input states are both $\ket{\tilde{\gamma}^{(1)}}$ from Eqn.~(\ref{initial_state}), then $\left|a_j\right|^2 = \left|a_j'\right|^2$ for all $j$.  In general, when the inputs satisfy $\left|a_j\right|^2 = \left|a_j'\right|^2$ for all $j$, we say the distillation is ``symmetric.''  The success probability is $P_{\mathrm{success}} = \sum_{y=0}^{N-1}\left|a_y\right|^4 \approx 0.67$, where the coefficients $\{a_j\}$ can be calculated using the method in Section~\ref{Fourier_analysis}.  The Fourier-basis weights of the output state are $\left|b_j\right|^2 = \left|a_j\right|^4/P_{\mathrm{success}}$.  The fidelity $F = \left|b_1\right|^2$ after one round of distillation is upper-bounded by $0.986$, so multiple rounds of distillation are needed to reach arbitrarily high fidelity.  This bounded fidelity also means that early rounds of distillation can use fewer than $n$ qubits to represent the intermediate Fourier states, as we explain in Section~\ref{fundamental_analysis}; before explaining that technique, we must quantify the fidelity in each round.

We define an $n$-qubit, $m$-round distilled $\ket{\tilde{\gamma}_m^{(1)}}$ Fourier state as having ``sufficiently high fidelity'' if its fidelity with the pure Fourier state satisfies
\begin{equation}
1-\left|\langle \tilde{\gamma}_m^{(1)} \mid \gamma^{(1)} \rangle \right|^2 \le \sin^2\left(\pi/2^n\right).
\label{fidelity_threshold}
\end{equation}
Subscript denotes how many rounds of symmetric distillation have been successfully applied, so initial state $\ket{\tilde{\gamma}^{(1)}} = \ket{\tilde{\gamma}_0^{(1)}}$.  In phase kickback, the constraint in Eqn.~(\ref{fidelity_threshold}) represents the highest accuracy that is needed.  When the $\ket{\tilde{\gamma}_m^{(1)}}$ register is used for phase kickback, there are two sources of error that we consider here.  The first is that the register $\ket{\tilde{\gamma}_m^{(1)}}$ is not pure, meaning it has non-zero overlap with some other Fourier basis state.  The second error source is that any phase rotation is truncated to $n$ bits of precision.  As a result, the truncated angle error is at most $\pi/2^n$ radians, which results in an upper bound on the rotation-gate error probability of $\sin^2\left(\pi/2^n\right) \approx \left(\pi/2^n\right)^2$.  In phase kickback using an $n$-qubit $\ket{\tilde{\gamma}_m^{(k)}}$ state, the combination of the two errors means that any resulting rotations are accurate to $\pm \pi/2^{(n-1)}$ radians, or at least $(n-1)$ bits.  Ultimately, $n$ is chosen based on the gate-accuracy requirements of the quantum algorithm.  We arbitrarily choose to balance the error from a noisy $\ket{\tilde{\gamma}_m^{(1)}}$ with the worst-case truncation-of-angle error.  If $\ket{\tilde{\gamma}_m^{(1)}}$ is used for other applications, such as complex-instruction-set quantum computing~\cite{Landahl2013}, a different accuracy may be required.

\begin{figure}
  \centering
  \includegraphics[width=\textwidth]{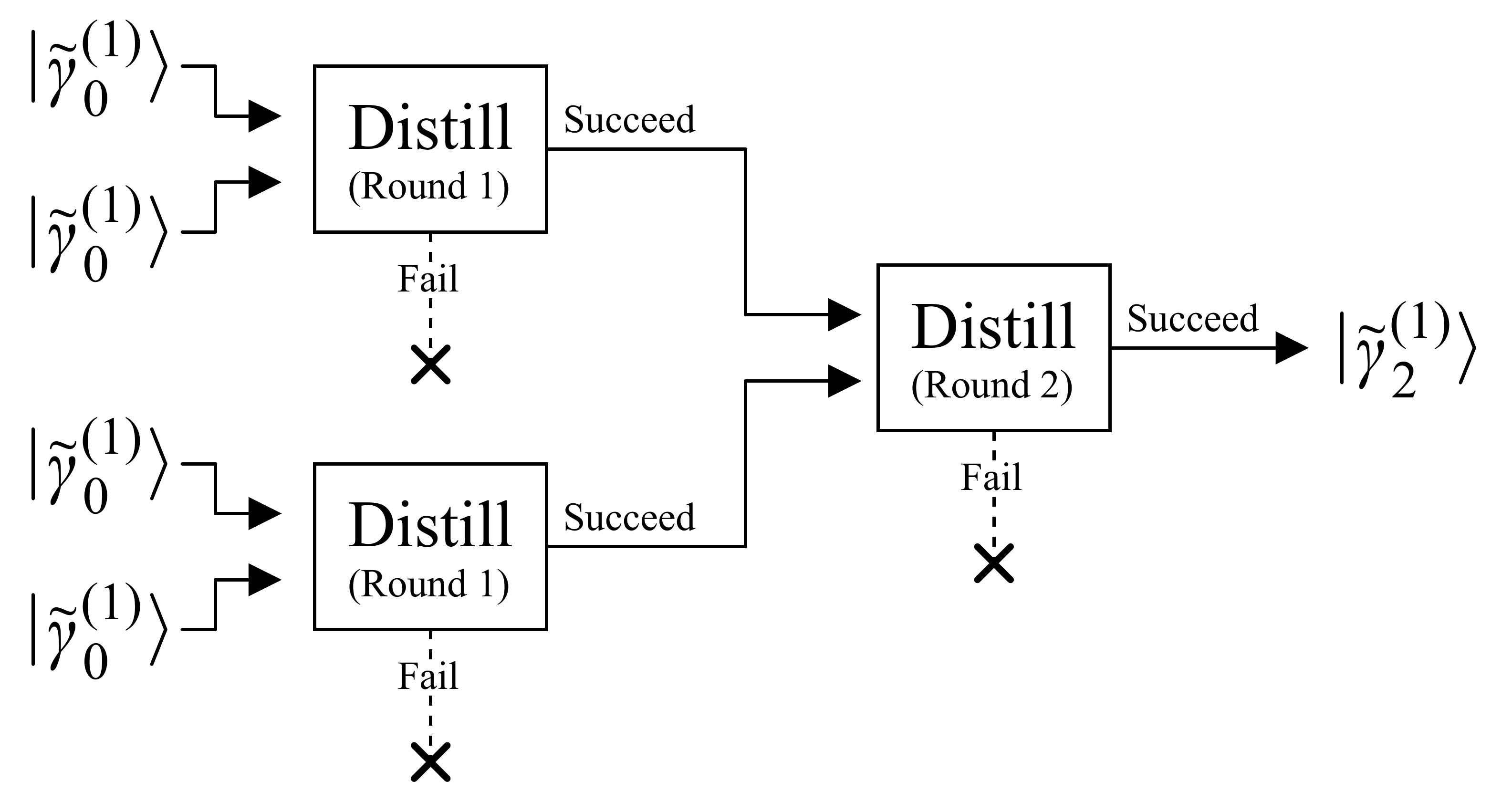}\\
  \caption{Symmetric distillation of Fourier states, where two rounds are shown.  The approximate initial states are defined in Eqn.~(\ref{initial_state}).  Each distillation step is implemented using the quantum circuit in Fig.~\ref{distillation_circuit}.  Distillation is probabilistic, and when a distillation circuit fails, that circuit and preceding steps that feed into it must be repeated.  The probability of failure decreases super-exponentially in round number, so the overhead of distillation failure is bounded.}
  \label{distillation_tree}
\end{figure}

The first round of distillation will produce a Fourier state accurate to about 5 bits.  To construct an $n$-qubit Fourier state, we develop a distillation protocol consisting of multiple rounds of symmetric distillation.  The symmetric distillation subroutines are arranged in a binary tree, as shown in Fig.~\ref{distillation_tree}.  Multiple low-fidelity $\ket{\tilde{\gamma}_0^{(1)}}$ input states are distilled to arrive at a single output state, $\ket{\tilde{\gamma}_r^{(1)}}$.  Subscript $r$ is the number of rounds of symmetric distillation, or the depth of this binary tree arrangement.  Choosing $r$ depends on the desired number of precision qubits $n$ in the Fourier state.  In the next section, we show that $r$ scales as $O(\log n)$.

\subsection{Fourier analysis and distillation efficiency}
\label{Fourier_analysis}
The distillation protocol can be understood by viewing probability amplitudes of the input state in the computational basis as discrete samples in time of a function $f(t) = e^{i 2 \pi \nu(t)}$ such that $\ket{\tilde{\gamma}_0^{(1)}} = \sum_{y=0}^{N-1} f(y/N)\ket{y}$.  In this picture, the probability amplitudes of the quantum state in the Fourier basis are related to Fourier-series coefficients $\{c_j\}$ given by
\begin{equation}
f(t) = \sum_{k = -\infty}^{\infty} c_j e^{i 2 \pi j t}.
\end{equation}
The correspondence exists because a quantum Fourier transform maps between computational and Fourier bases.  The number of ``samples'' is $N = 2^n$, the number of computational basis states.  Discrete sampling causes aliasing according to the Shannon-Nyquist Theorem, so Fourier-basis probability amplitudes $\{a_j\}$ are related to the Fourier series of $f(t)$ by
\begin{equation}
a_j = \sum_{x = -\infty}^{\infty} c_{(Nx + j)}.
\label{coefficient_relationship}
\end{equation}
If $N$ is sufficiently large (\emph{e.g.} $n \ge 6$), then $a_j \approx c_j$ for $0 \le j < N/2$ or $a_{(N+j)} \approx c_j$ for $-N/2 \le j < 0$, because the coefficients $c_j$ decay in magnitude asymptotically as $\left|c_j\right|^2 \propto 1/j^2$, which means the error from neglecting aliased frequencies is suppressed exponentially in $n$.  This asymptotic upper bound follows from Parseval's theorem for any signal that is square-integrable over its period, a condition which corresponds to normalized quantum states.

\begin{figure}
  \centering
  \includegraphics[width=\textwidth]{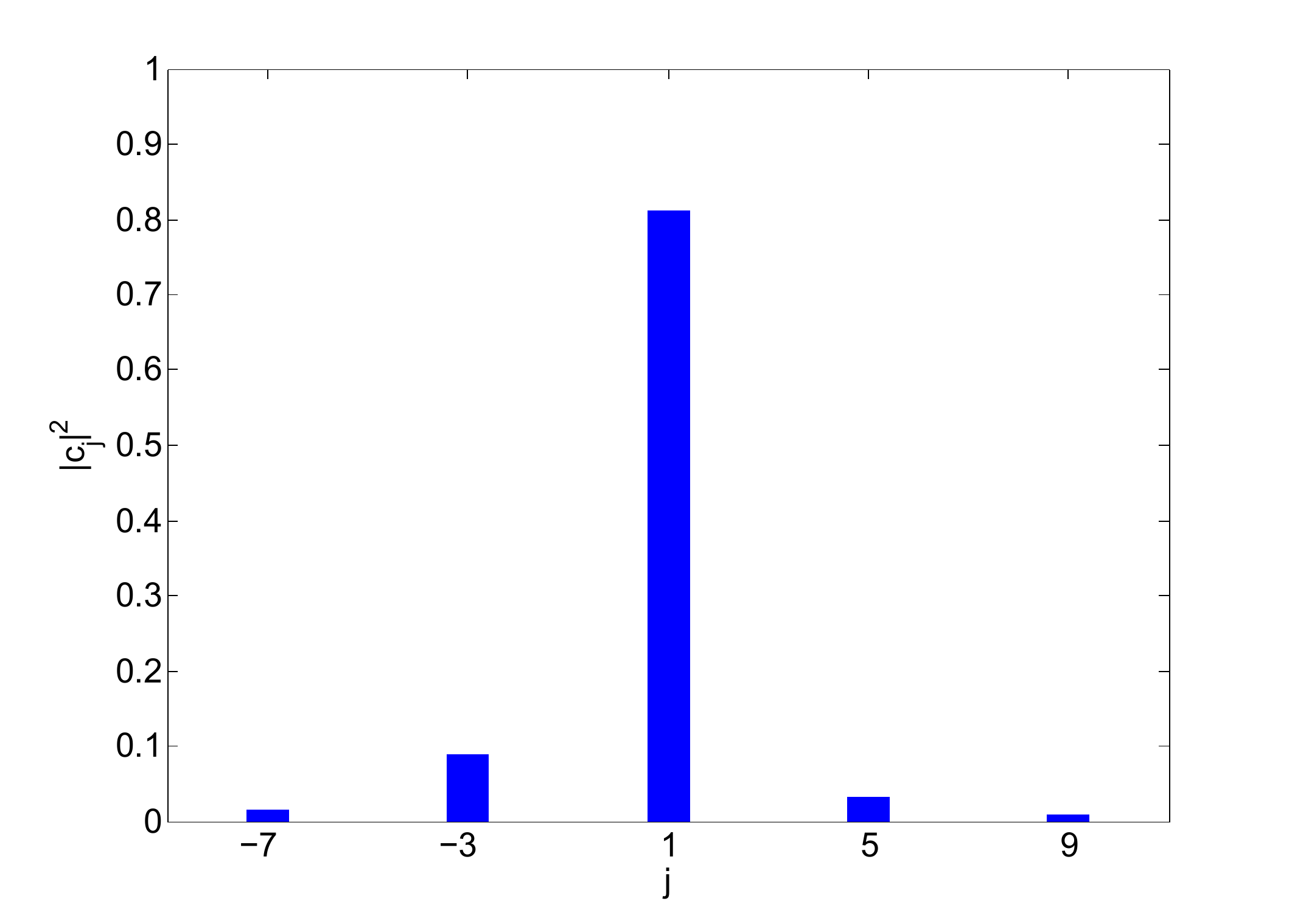}\\
  \caption{Frequency spectrum of $f(t)$.  The height of a bar at $j$ corresponds to the squared magnitude $|c_j|^2$, where $\{c_j\}$ are Fourier-series coefficients of $f(t)$.  Eqn.~(\ref{coefficient_relationship}) relates these series coefficients to Fourier-basis amplitudes.}
  \label{spectrum}
\end{figure}

In each approximate $\ket{\tilde{\gamma}_0^{(1)}}$ register, the first qubit is the most significant bit in a binary encoding of equally-spaced time coordinates for samples of $f(t)$.  By using only $Z$ and $S$ rotations, we are effectively discretizing the phase of $f(t) = e^{i 2 \pi \nu(t)}$ to two bits of precision as a piecewise-constant function over four equally-sized intervals in the domain $t \in [0,1)$.  We can readily calculate the $j^{\mathrm{th}}$ Fourier series coefficient of this function:
\begin{eqnarray}
c_j & = & \int_{0}^{1} e^{i 2 \pi (\nu(t) - jt)} dt \nonumber \\
    & = & \sum_{r = 1}^4 \int_{(m-1)/4}^{m/4} e^{-i 2 \pi (jt - (r-1)/4)} dt \nonumber \\
    & = & \left(\frac{1-i}{2 \pi j}\right) \sum_{m=1}^4 e^{-i \pi m (j-1) /2},
\end{eqnarray}
which is valid everywhere except $j=0$, in which case $c_0 = 0$.  Sign convention follows Eqn.~(\ref{define_Fourier_state}).  The only nonzero terms occur for $j \equiv 1 \; (\mathrm{mod} \; 4)$, and they are $c_j = (2-2i)/(\pi j)$.  The squared magnitudes of the largest Fourier components for state $\ket{\tilde{\gamma}_0^{(1)}}$ are plotted in the spectrum in Fig.~\ref{spectrum}.  The expression for Fourier-series coefficients allows us to derive bounds on distillation performance and hence the necessary number of rounds of distillation.  For example, since there is no relative phase between these coefficients, initial Fourier states have $|a_1|^2 > |c_1|^2$, where $|c_1|^2 = 8/\pi^2$.

The output state of the distillation protocol, conditioned on success, has modified Fourier components.  In symmetric distillation, the relative magnitude of each component to the fundamental harmonic is squared.  After normalization, the largest component at $j = 1$ is amplified, while the rest are suppressed.  The second-largest component is the ``sideband'' at $j = -3$ in Fig.~\ref{spectrum}.  Note that because of aliasing in the frequency spectrum, $c_{(-3)}$ maps to $a_{(N - 3)}$; we assume that Fourier-series term $c_{(N - 3)}$ is negligibly small.

Distillation proceeds until the sidebands are suppressed to a sufficiently low level.  The rate at which these sidebands are suppressed dictates how many rounds of distillation are required, which determines the total number of gates in the protocol.  This rate is limited by the ratio in magnitudes between the fundamental $j = 1$ harmonic and the second-largest sideband at $j = -3$.  This behavior is analogous to the rate of convergence in Markov-chain Monte Carlo, which depends on the magnitude of the second-largest eigenvalue of the state transition matrix (the largest eigenvalue of a stochastic matrix is 1).  Successful symmetric state distillation through $r$ rounds modifies each Fourier-basis amplitude from $a_j$ to $b_j^{(r)}$ according to
\begin{equation}
\left|b_j^{(r)}\right|^2 = \frac{1}{C}\left(\left|a_j\right|^2\right)^{2^{r}},
\end{equation}
where
\begin{equation}
C = \sum_{j=0}^{N-1} \left(\left|a_j\right|^2\right)^{2^{r}}
\end{equation}
is the normalization.  Since any sidebands to the fundamental harmonic ($j = 1$ in this case) will be suppressed super-exponentially in $r$, we need to only focus on the magnitude of the largest sideband at $c_{(-3)}$, which will dominate the error in the output of distillation.  As a result, the error $\epsilon = 1 - \left|\left\langle \tilde{\gamma}_m^{(k)} \mid \gamma^{(k)} \right\rangle \right|^2$ in the distilled Fourier state is closely approximated by
\begin{equation}
\epsilon \approx \left(\left|c_{(-3)}\right|^2/\left|c_1\right|^2\right)^{2^{r}}.
\label{distillation_error}
\end{equation}
Consequently, the ratio $\left|c_{(-3)}\right|^2/\left|c_1\right|^2 = 9$ (exactly) dictates how fast error is suppressed through distillation.  Since we require that $\epsilon \le \left(\pi/2^n\right)^2$, we can determine the number of rounds of distillation as
\begin{equation}
R = \left\lceil \log_2 \left(\frac{2n - 2 \log_2 \pi}{\log_2(|c_1|^2/|c_{(-3)}|^2)}\right) \right\rceil.
\label{num_rounds}
\end{equation}
This expression can be simplified to $R = \left\lceil \log_2(0.63n - 1.04) \right\rceil$ (approximately), which shows that $R$ scales as $O(\log n)$.  Moreover, Eqn.~(\ref{distillation_error}) shows that the error at the output of each successive round of distillation is squared.  Eqn.~(\ref{fidelity_threshold}) shows that the number of qubits needed to represent an approximate Fourier state is $O(\log \epsilon)$, so the smallest allowable size in qubits of intermediate distilled states will double after each round.  The next section uses this technique to save resources.

\section{Resource analysis for distilling the fundamental Fourier state}
\label{fundamental_analysis}
This section shows that distilling the fundamental $n$-qubit Fourier state is efficient in the sense that it requires at most $O(n \log n)$ Toffoli gates and total gates, with circuit width $2n + O(1)$ qubits.  After each round of distillation, the number of bits of precision in the Fourier states doubles, so we also double the number of qubits going into the next round.  The procedure is: (a)~after one round of distillation, each Fourier state is accurate to $s$ qubits; (b)~append $s$ more qubits in the $\ket{+}$ state to each register; (c)~repeat distillation on the input states of size $2s$ qubits.  The additional error of making a larger approximate Fourier state by appending $s$ qubits in the $\ket{+}$ state is less than the error already present, so the fidelity is not reduced appreciably.  The extra qubits provide space for the output state of distillation (if it succeeds) to contain twice as many accurate qubits.

\begin{figure}
  \centering
  \includegraphics[width=\textwidth]{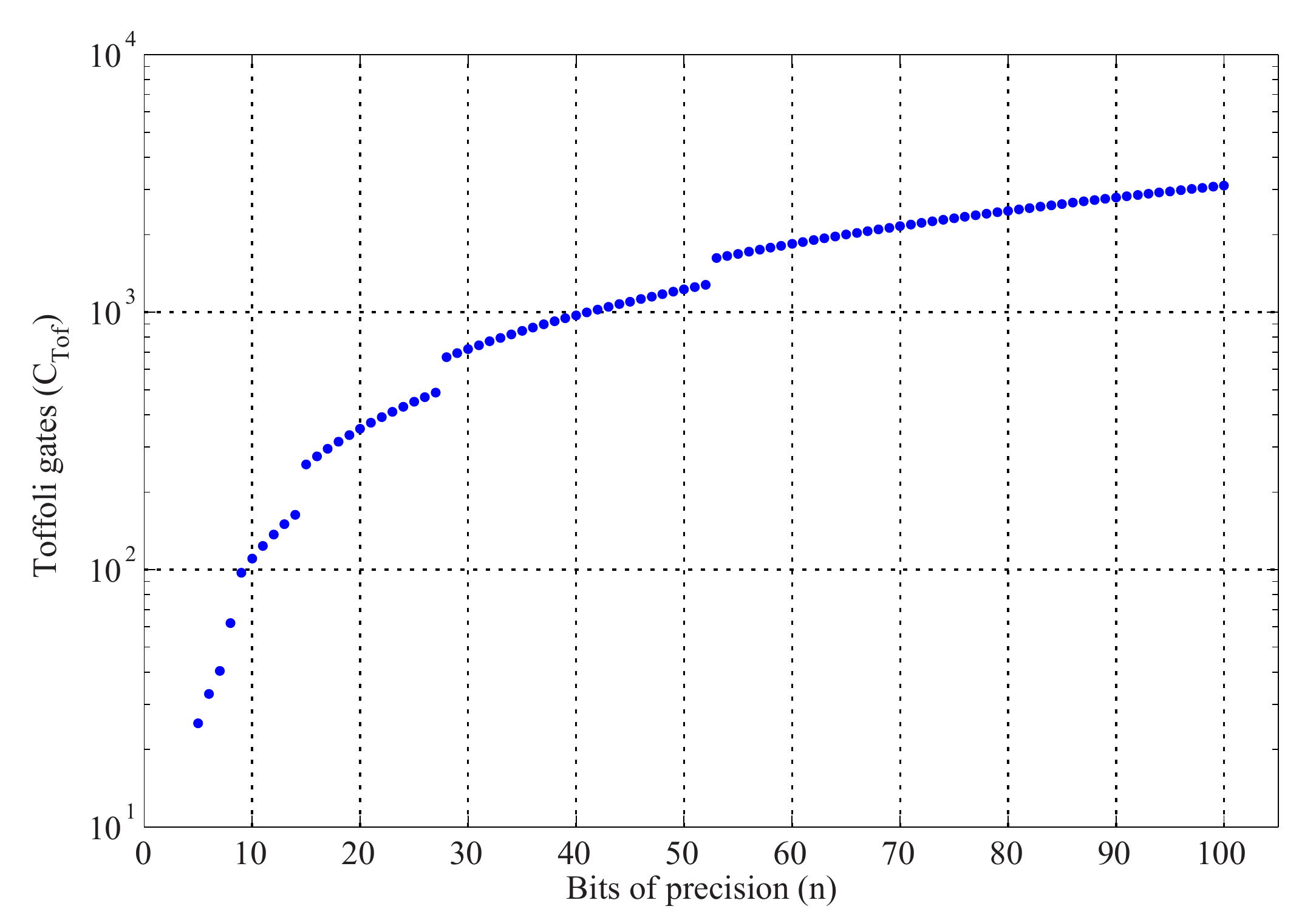}\\
  \caption{Expected number of Toffoli gates consumed in the distillation of an $n$-qubit Fourier state.  Uncertainty in number of Toffoli gates is a result of distillation success being probabilistic in each round.  No other non-Clifford gates are required, and the Toffoli gates dominate the cost of the adder circuits within distillation.}
  \label{Toffoli_cost}
\end{figure}

Each round of distillation on $s$-qubit registers uses addition circuits that each require $(2s-4)$ Toffoli gates~\cite{Cuccaro2004} (note that the carry-out Toffoli is unnecessary and removed).  If there are $R$ rounds of distillation, then the $r^{\mathrm{th}}$ round requires $2^{(R-r)}$ adder circuits.  If we begin with $s$ qubits per Fourier state going into the first round and double the number of qubits in each subsequent round, then the total number of Toffoli gates in the entire distillation protocol is
\begin{equation}
C_{\mathrm{Tof}} = \sum_{r = 1}^R 2^{(R-r)} (2^{(r+1)}s-4) = 2^{(R+1)} R s - 2^{(R+2)} + 4.
\end{equation}
Since $R$ scales as $O(\log n)$, then $C_{\mathrm{Tof}}$ scales as $O(n \log n)$.  We use $s = 5$ since the output of the first round is accurate to about 5 bits of precision.  Eqn.~(\ref{num_rounds}) gives an exact expression for $R$.  Fig.~\ref{Toffoli_cost} plots the number of Toffoli gates required for distillation up to $n = 100$ bits of precision, which is the most precision one could imagine needing for a quantum algorithm.  For example, 10 bits of precision is more than sufficient for 4096-bit Shor's algorithm~\cite{Fowler2004}.  In this case the approximately 100 Toffoli gates needed to distill $\ket{\gamma^{(1)}}$ are negligible in comparison to the rest of the algorithm~\cite{Jones2012}.  We emphasize that even if multiple copies of a Fourier state are required, the distillation need only be performed once.  Fourier states of size $n$ qubits can be cloned using a single adder circuit requiring $(2n-4)$ Toffoli gates.

The last round of distillation uses $n + O(1)$ qubits for each input register, which may be less than $2^{(R-1)}s$.  The additive constant appears because one might need to distill to $(n+1)$ or $(n+2)$ qubits in the final output, compensating for errors introduced by truncating the size of Fourier states in earlier rounds.  Each round of symmetric distillation doubles the number of qubits per Fourier state, but the number of Fourier states is reduced by half.  Consequently, the circuit width of the protocol is at most $2n + O(1)$ qubits, because the final round uses two registers of size $n + O(1)$ qubits.

\section{Resource comparison for approximating rotation gates}
\label{resource_comparison}
One of the main reasons to distill Fourier states is that they facilitate fault-tolerant phase rotations with phase kickback~\cite{Jones2012b}, so we determine the resource costs of this method and compare it to alternatives.  Approximating an arbitrary phase rotation with error probability $\sin^2\left(\pi/2^{n-1}\right)$ requires an $n$-qubit, distilled $\ket{\gamma^{(1)}}$ Fourier state, including the residual error probability from distillation.  As shown previously, preparing such a state requires $O(n \log n)$ Toffoli gates, but this initialization need only be performed once.  Each phase rotation uses an addition circuit, which requires at most $2n-4$ Toffoli gates~\cite{Cuccaro2004}; however, one of the addends is a known value, so we can simplify the adder and ``short-circuit'' half of the Toffoli gates, replacing them with Clifford gates.  In this special case, a single-qubit phase rotation with a precision of $\pi/2^{n-1}$ radians, which is $(n-1)$ bits of precision, requires just $(n-2)$ Toffoli gates, $n$ qubits for the Fourier states, and $(n-1)$ ancilla qubits for the internal carry operations of the adder.  Forming controlled-rotation gates is simple as well.  Each additional control input to the multi-qubit gate requires one more Toffoli gate and one more ancilla qubit.

The best results for approximating single-qubit rotations with sequences of $T$~gates and Clifford gates each require about 3 $T$~gates per bit of precision~\cite{Fowler2009,Kliuchnikov2012}.  These methods require at most 2 ancilla qubits instead of $(2n-1)$ for phase kickback, but the total resource costs from non-Clifford gates is higher.  Reference~\cite{Kliuchnikov2012} estimates the number of $T$~gates is $C_T = 3.21\log_2(1/\epsilon_{\mathrm{F}}) - 6.93$; gate-error $\epsilon_{\mathrm{F}}$ is defined in Ref.~\cite{Fowler2009} as
\begin{equation}
\epsilon_{\mathrm{F}} = \sqrt{1-\frac{1}{2}\left|\mathrm{tr}(U^{\dag}\tilde{U})\right|},
\end{equation}
where $\tilde{U}$ is the fault-tolerant sequence approximating gate $U$.  In the case of a phase kickback rotation accurate to $p$ bits, the error would be
\begin{equation}
\epsilon_{\mathrm{F}} = \sqrt{1-\frac{1}{2}\left|1+\exp(i\pi/2^p)\right|} \approx \frac{1}{\sqrt{8}}\left(\frac{\pi}{2^p}\right),
\end{equation}
where approximation is correct to at least four significant figures for $p \ge 6$.  A phase kickback rotation accurate to $p$ bits has the succinct error expression $\log_2(1/\epsilon_{\mathrm{F}}) \approx p - 0.15$.

We give direct comparison of resource costs in the two methods approximating rotation gates.  To construction a rotation gate accurate to $p$ bits, one would require (in terms of non-Clifford gates):
\begin{itemize}
\item $(3.21p - 6.45)$ $T$~gates (on average) using an approximation sequence~\cite{Kliuchnikov2012}, or
\item $(p-1)$ Toffoli gates using phase kickback~\cite{Jones2012b}.
\end{itemize}
One could produce a Toffoli gate using 4 $T$~gates~\cite{Jones2013,Eastin2012}, in which case approximation sequences and phase kickback have similar costs in terms of non-Clifford gates.  However, efficient fault-tolerant constructions have been found for the Toffoli gate which further reduce the required resources to about the cost of a single $T$~gate, or even lower~\cite{Jones2013}.  Consequently, the total cost of non-Clifford gates is expected to be lower when using phase kickback.  Given that the cost of either $T$ or Toffoli is much greater than a Clifford gate or ancilla qubit (by about two orders of magnitude~\cite{Fowler2012,Fowler2013}), there is considerable evidence that phase kickback is the current lowest-overhead construction for a fault-tolerant phase rotation.  Further work is needed to develop explicit expressions for resource costs as a function of hardware parameters and the methods of error correction selected.

\section{Distilling arbitrary Fourier states}
\label{arbitrary_distillation}
We outline here a procedure for distilling any $\ket{\gamma^{(k)}}$ Fourier state, leaving the detailed analysis for future work.  Arbitrary values of $k$ are need for QVR phase kickback~\cite{Jones2012b}, which is useful in quantum simulation and some implementations of the linear-systems algorithm algorithms~\cite{Jones2012b,Clader2013}.  Ref.~\cite{Jones2012b} gives a method for transforming any $n$-qubit $\ket{\gamma^{(k)}}$ with odd $k$ to any other $\ket{\gamma^{(k')}}$, using $O(n^2)$ gates.  Since it requires phase kickback with successively larger addition circuits, the number of Toffoli gates is
\begin{equation}
\sum_{s = 3}^{n-1}(s-2) = \frac{(n-3)(n-2)}{2}.
\end{equation}
That protocol is deterministic and does not require distillation, assuming one already has a Fourier state.  The protocol in Sec.~\ref{protocol_section} could distill $\ket{\gamma^{(1)}}$, which could then be transformed into any $\ket{\gamma^{(k)}}$.

Approximations of any $\ket{\gamma^{(k)}}$ can also be distilled using the protocol in Section~\ref{protocol_section}.  To have reasonably good efficiency, the approximate initial states need to have substantial fidelity with the desired pure state, say $F > 0.5$.  There are at least two possible approaches.  One is to start $\ket{\gamma^{(0)}} = \ket{+}^{\otimes n}$, then apply QVR for each `1' bit in the binary representation of $k$, using a $\ket{\gamma^{(1)}}$ state truncated to $O(\log n)$ qubits.  Each QVR gate need only be accurate to error $O(2^{-\log n}) = O(1/n)$, as there are at most $n$ such operations, so the aggregate error is of order unity, meaning we can bound it below 0.5.  These approximate states are then distilled, but they are all of size $n$ qubits, so the total number of Toffoli gates is $O(n^2)$.  Whether this method is more efficient than the deterministic construction is not yet clear.  A second way to prepare approximate states is to split the quantum register encoding a desired Fourier state into two registers of roughly equal size, and first prepare these approximately through distillation.  This method can be applied recursively until the input states are small, like the $s=5$ starting state above.  For the protocol in Section~\ref{fundamental_analysis}, many of the intermediate registers are $\ket{\gamma^{(0)}}$, which can be constructed with Clifford gates; this will not always hold for arbitrary $k$, so the number of Toffoli gates needed here is higher than the distillation of $\ket{\gamma^{(1)}}$ and may also be $O(n^2)$ for this method.  Since Fourier states used for QVR phase kickback can be reused, the number of \emph{different} values of $k$ used by an algorithm will dictate whether seeking optimized state-preparation protocols is a worthwhile endeavor.

\section{Discussion}
Although we only use $Z$ and $S$ gates to initialize approximate Fourier states, one could also use smaller-angle magic states $R_z(\pi/2^x)\ket{+}$ for $x > 2$, which would increase success probability and decrease the number of rounds.  However, this approach would require distillation of those small-angle magic states or approximation of the small-angle rotations~\cite{Landahl2013}.  Adding a low-fidelity $T$~gate to produce a more accurate initial state might be advantageous, but dramatic improvements using smaller-angle rotations are not expected for typical quantum-computing parameters.  Conversely, Fourier states would readily enable the complex-instruction-set computing of Ref.~\cite{Landahl2013}, because each Fourier state is the tensor product of the desired small-angle magic states.  Fourier states can be cloned with just $(2n-4)$ Toffoli gates using QVR phase kickback~\cite{Jones2012b}.  As such, phase kickback is a better way to produce these magic states than distilling them individually.

Since recent circuit constructions have substantially lowered the cost of a fault-tolerant Toffoli gate~\cite{Jones2013,Eastin2012}, one should consider whether further improvements to $T$~gates are possible.  Reductions in $T$-gate cost could make approximation sequences a better choice than phase kickback for approximating gates, and this comparison determines whether our results in Fourier-state distillation are useful.  We give two arguments for why substantial lowering of $T$-gate costs is unlikely.  First, magic-state distillation for $T$~gates places considerable constraints on the underlying quantum codes.  It has been conjectured that a protocol distilling states from error $\epsilon$ to $O(\epsilon^d)$ yields an output/input distillation fraction less than $1/d$~\cite{Bravyi2012}, and the only known codes that approach this limit are so complicated as to make implementation impractical~\cite{Jones2012c}.  From this we infer that magic-state distillation for $T$~gates cannot improve substantially unless a completely different method is discovered.  Second, error correction for Toffoli gates is much more efficient by comparison, and there is further room for improvement.  In the best Toffoli constructions, one uses $T$~gates (at lower fidelity though perhaps distilled) inside an error-detecting circuit to suppress gate errors; with this approach, logical Toffoli is the only non-Clifford gate visible to the algorithm.  If the conjecture above were extended to resource costs for Toffoli gates, then the protocol in Refs.~\cite{Jones2013,Eastin2012} saturates the limit through one round, which is why a Toffoli gate costs about the same resources as a $T$~gate with current methods; further improvements are expected to reduce the cost of a logical Toffoli gate below that of a logical $T$~gate.

Recent work suggests ``V-basis'' rotations may also be efficient at approximating arbitrary rotation gates~\cite{Duclos2012,Bocharov2013}.  These non-Clifford gates can be generated from a distillation procedure that uses $T$~gates~\cite{Duclos2012}.  Further work is needed to determine the fault-tolerant resource costs of V-basis methods and how these costs compare to those of phase kickback or other methods.

By demonstrating how to distill Fourier states and hence complete the phase kickback protocol, this paper has two broader implications. (1)~Quantum computing with only Clifford and Toffoli gates is efficient, and simple constructions for arbitrary gates are known. (2)~Quantum algorithms using an instruction set of (Clifford+Toffoli) gates could require fewer resources than those using an instruction set of (Clifford+$T$) gates, so the Toffoli gate deserves serious attention as the non-Clifford operation in fault-tolerant quantum computing.


\end{document}